\documentclass{ws-procs9x6}
\newcommand{\be}{\begin{equation}}
\newcommand{\ee}{\end{equation}}
\newcommand{\bea}{\begin{eqnarray}}
\newcommand{\eea}{\end{eqnarray}}
\newcommand{\nn}{\nonumber}

\begin{document}

\title{NEW OPEN AND HIDDEN CHARM SPECTROSCOPY}

\author{P. COLANGELO$^*$, F. DE FAZIO,  R. FERRANDES,  S. NICOTRI}

\address{Istituto Nazionale di Fisica Nucleare, Sezione di Bari,\\
Bari, Italy
\\$^*$E-mail: pietro.colangelo@ba.infn.it}

\begin{abstract}
Many new results on open and hidden charm spectroscopy have been obtained recently. We present
a short review of the experimental findings in the meson sector, of the theoretical interpretations and of the open problems, with a discussion  on  the possibility that some mesons are not quark-antiquark  states.
\end{abstract}


\bodymatter

\section{Introduction}\label{sec:intro}
Observation of a long list of  new hadrons has been recently  reported by experiments  at   $e^+ e^-$ and $p \bar p$ colliders,   by fixed target experiments  and  by reanalyses of old data.  We can use Leporello's words in  Mozart's  Don Giovanni:  {\it Madamina, il catalogo \`e questo:}
\footnote{My lady, this is the list:} 

 $D_{sJ}^*(2317)$, 
 $D_{sJ}(2460)$,  $D_{sJ}(2632)$,  $D_{sJ}(2860)$,  $D_{0}^*(2308)$,
 $D_{1}^\prime(2440)$,    $h_c$,   $\eta_c^\prime$,  $X(3872)$,  $X(3940)$,   $Y(3940)$, $Z(3930)$,   $Y(4260)$,  $\Upsilon(1D)$,   $B_{1}$,  $B_{2}$,  $B_{s2}$,  $\Theta(1540)^+$,  $\Theta_c(3099)$,  $\Xi_{cc}(3518)$,  \dots  

Not all the states in the list  have been  confirmed 
($D_{sJ}(2632)$, $\Theta(1540)^+$,  $\Theta_c(3099)$) and therefore we can ignore them.  Other states ($\Xi_{cc}(3518)$) are baryons, deserving a dedicated analysis, and mesons with open ($B_{1}$,  $B_{2}$, $B_{s2}$)  or  hidden beauty ($\Upsilon(1D)$),  that  we do not discuss  here.  
 
We  only  consider  mesons with open and hidden charm.  The wealth of information collected in recent years is impressive:  not only  the number of known states has nearly doubled, but a few experimental observations seem to challenge the current  picture of mesons as simple quark-antiquark configurations.  Therefore, it is important to  search  the  signatures 
 allowing us to assign a given state to a particular multiplet, so that the hints of exotic structures can be
 clearly interpreted.  The next Sections are devoted to such a discussion, considering separately  the case of open charm mesons, which at present can be classified according to known rules, and that of hidden charm states where a couple of  mesons seem to escape simple classification schemes.
 \footnote{For other recent reviews on  this subject  see Refs.~\refcite{Colangelo:2003vg,swanson}.}
\section{Mesons with open charm}\label{sec:openc}
In QCD, for hadrons containing a single heavy quark Q and in the limit $m_Q \to \infty$, 
there is a spin-flavour symmetry  due to the decoupling of the heavy quark from the dynamics
of the light degrees of freedom (light quarks and gluons). Therefore,
 it is possible to classify states containing the heavy quark $Q$ according to the total angular
momentum  $s_\ell$ of the  light degrees of freedom. For mesons, 
states belonging to doublets with the same
$s_\ell=s_{\bar q}+ \ell$, with $s_{\bar q}$ the spin of the light antiquark  and $\ell$ the orbital angular
momentum  relative to the heavy quark, are degenerate in mass in the large $m_Q$ limit. In case of
charm, $D^{0,+}$  and  $D^{*0,+}$, $D_s$  and  $D^*_s$ are the states in
the  $s_\ell^P=\frac{1}{2}^-$  $c \bar u (\bar d, \bar s)$ doublet, corresponding  to  $\ell=0$. The mass difference between the members of the doublet is $O(\frac{1}{m_c})$, and vanishes when
$m_c  \to \infty$. 

For $\ell=1$ there are two doublets  with  $s_\ell^P=\frac{1}{2}^+$ and  $s_\ell^P=\frac{3}{2}^+$, 
for $\ell=2$ two other doublets with $s_\ell^P=\frac{3}{2}^-$ and  $s_\ell^P=\frac{5}{2}^-$, and so on.
The spin-flavour symmetry is important not only for spectroscopy, but also  
  for the classification of strong decay modes and for evaluating the rates, since
decays involving heavy mesons belonging to the same doublets are related.
For example, the decays
of mesons belonging to the  $s_\ell^P=\frac{3}{2}^+$ doublet in one  light  pseudoscalar and one heavy
$s_\ell^P=\frac{1}{2}^-$ meson occur in $d-$wave, so that these states are expected, {\it ceteris paribus}, to be narrower than the states belonging to the doublet  $s_\ell^P=\frac{1}{2}^+$, which decay to the same final states by $s-$wave transitions. These observations  are at the basis of  the analyses of the new mesons observed in $c \bar q$ and in $c \bar s$ systems. They must be used 
together with the consideration  that  $\frac{1}{m_Q}$ effects  can be important in  case of charm:
for example, the two $1^+$ states belonging to $s_\ell^P=\frac{1}{2}^+$ and $\frac{3}{2}^+$ doublets,
 due to the finite charm quark mass, could mix with a mixing angle $\theta_c$  to provide the physical axial vector mesons. Such  effects must  be investigated on the basis of  the experimental observation.
\subsection{$c \bar q$ mesons: $D_{0}^*(2308)$ and $D_{1}^\prime(2440)$.}
Information about  broad $c \bar q$ mesons, one scalar and one axial vector charmed meson that can be interpreted as the states belonging to the  $s_\ell^P=\frac{1}{2}^{+}$ $c\overline{u}$,  $c\overline{d}$  doublets,  comes from  Cleo\cite{cleobroad},  Belle\cite{Bellelarghi} and  Focus\cite{focus} Collaborations. 
The resonance parameters are reported in  Table~\ref{tab:statilarghi}; they are obtained observing  that the $D \pi$ and $D^*\pi$ mass distributions,  produced for example in $B \rightarrow D^{**}\pi$ with a $D^{**}$  a generic $\ell=1$ meson, require  contributions with scalar or axial vector quantum numbers.
Improved  determinations of mass and width of the two other
positive parity charmed states $D_1$ $(J^P=1^+$)  and $D_2$ $(J^P=2^+$) have been
obtained, together with a measurement of the mixing
angle between the two $1^+$ states. It  is  small: $\theta_c=-0.10 \pm 0.03 \pm
0.02 \pm 0.02$~rad ($\simeq -6^0$)  \cite{Bellelarghi}.
%
%%%%%%%%%%%%%%%%%%%%%%%%%%%%%%%%%%%%%%%%%%%%
\begin{table}
\tbl{Mass and width  of broad resonances observed in D$\pi$ and
D$^*\pi$.}
{\begin{tabular}{cccc}
      \hline
      \ & \ &  Belle\cite{Bellelarghi} &   Focus\cite{focus}  \\
      \hline
      \ $D_0^{*0}$ & \begin{tabular}{c} M (MeV) \\ $\Gamma$ (MeV)
\end{tabular}   &
      \begin{tabular}{c} $2308\pm17\pm15\pm28$ \\ $276\pm21\pm18\pm60$
\end{tabular}   &
      \begin{tabular}{c}  $2407\pm21\pm35$ \\ $240\pm55\pm59$ \end{tabular} 
\\ \hline
         \ $D_0^{*+}$ & \begin{tabular}{c} M (MeV) \\ $\Gamma$ (MeV)
\end{tabular}   &
      \begin{tabular}{c} $$ \\ $$
\end{tabular}   &
      \begin{tabular}{c}  $2403\pm14\pm35$ \\ $283\pm24\pm34$ \end{tabular} 
\\
      \hline
      \ & \ &  Belle\cite{Bellelarghi} &   Cleo\cite{cleobroad}\\
      \hline
      \ $D_1'^{0}$ & \begin{tabular}{c} M (MeV) \\ $\Gamma$ (MeV)
\end{tabular}   &
      \begin{tabular}{c} $2427\pm26\pm20\pm15$\\
                                       $384_{-75}^{+107}\pm24\pm70$\end{tabular}
      &\begin{tabular}{c} $2461^{+41}_{-34}\pm10\pm32$\\
                                       $290_{-79}^{+101}\pm26\pm36$\end{tabular}
                                       \\
      \hline
    \end{tabular}}\label{tab:statilarghi}
\end{table}

\subsection{$c \bar s$ mesons: $D_{sJ}^*(2317)$, 
 $D_{sJ}(2460)$ and $D_{sJ}(2860)$.}
\begin{figure}[tbhp]
\begin{center}
\begin{tabular}{cc}
\epsfig{file=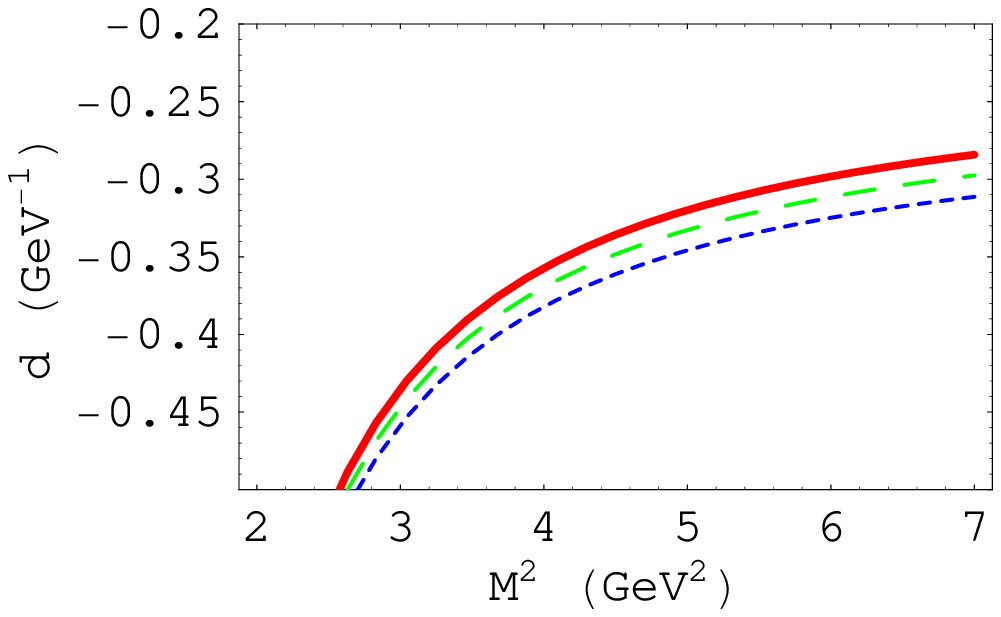, width=0.4\textwidth}&\epsfig{file=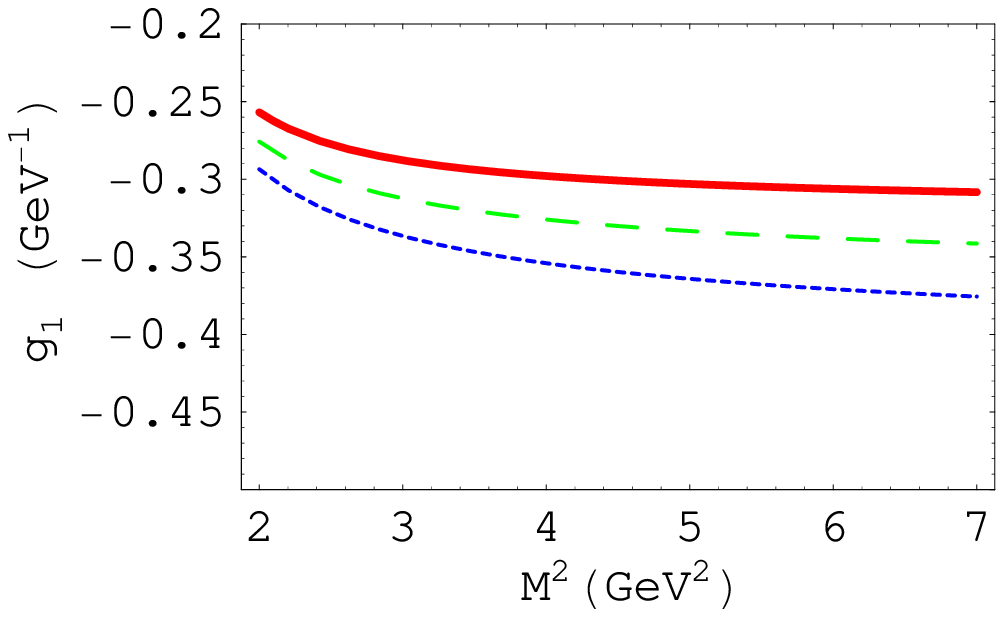, width=0.4\textwidth}\\
\epsfig{file=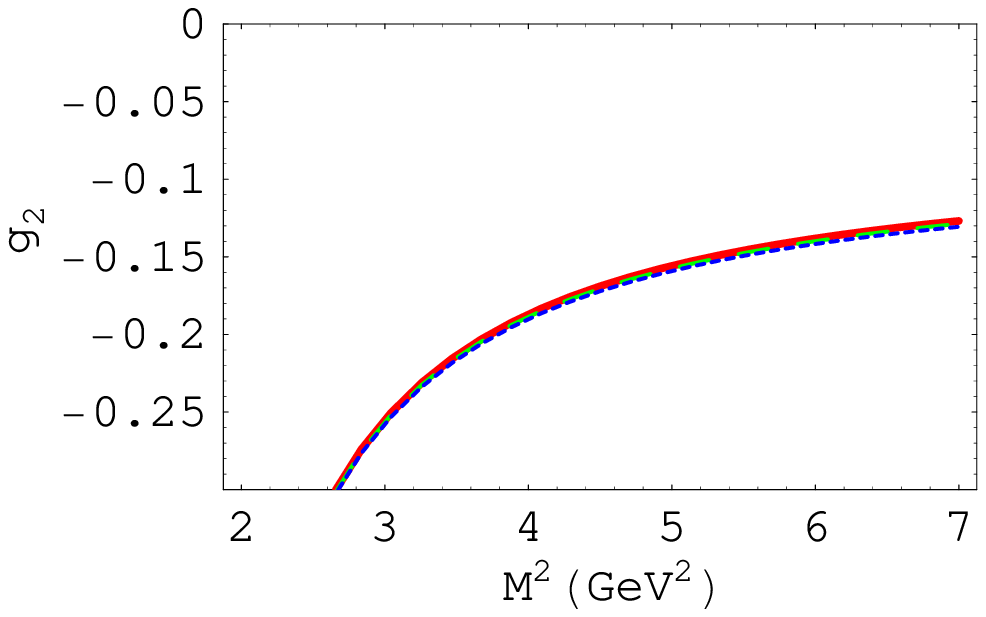, width=0.4\textwidth}& \epsfig{file=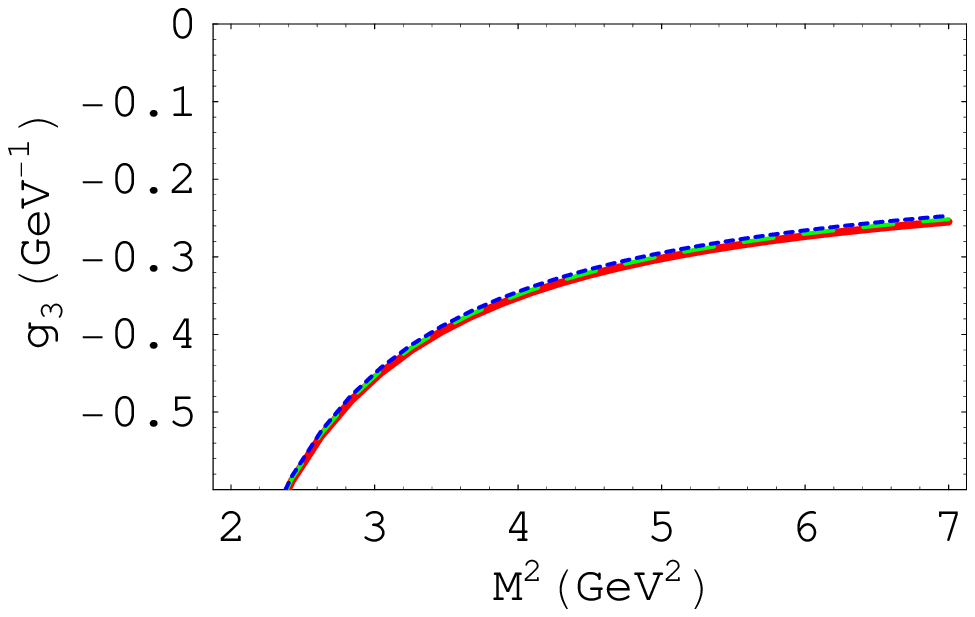, width=0.4\textwidth}\\
\end{tabular}
\end{center}
\caption{Results for the hadronic parameters in 
eq.(\ref{eq:ampDs1b}); $M^2$ is the Borel parameter.}
\label{fig:results}
\end{figure}

 $D_{sJ}^*(2317)$ and  $D_{sJ}(2460)$ were found at the $B$ factories in $D_s \pi^0$ and $D_s^* \pi^0, D_s^* \gamma$  distributions,  respectively, in $e^+e^-$ continuum  and in $B$ decays\cite{Aubert:2003fg}. Their 
widths are unresolved,  and this has arisen  doubts about their  identification as the scalar and axial vector $s_\ell^P= \frac{1}{2}^+$ $c  \bar s$ mesons  ($D_{s0}$ and $D_{s1}^\prime$), forming, together with $D_{s1}(2536)$ and  $D_{s2}(2573)$, the set of   four low-lying $\ell=1$  states.
However,  the masses of $D_{sJ}^*(2317)$ and $D_{sJ}(2460)$ are below their respective thresholds for strong decays, $DK$ and $D^* K$, therefore the small width is natural.  Moreover,
analyses of radiative transitions, that probe the structure of hadrons, support  the $c\bar s$ interpretation
 of the two states\cite{Godfrey:2003kg,colangelo1}.  For example, by  Light-Cone QCD sum rules one can compute the hadronic parameters $d, g_1, g_2$ and $g_3$  governing the $D^*_{sJ}(2317) \to D_s^* \gamma$ and  $D_{sJ}(2460) \to D_s^{(*)} \gamma, \,  D_{sJ}^*(2317) \gamma$ decay amplitudes\cite{Colangelo:2005hv}:
\bea
\langle \gamma(q,\lambda) D_s^*(p,\lambda^\prime)| D_{s0}(p+q)\rangle &=&  e d  \left[ (\varepsilon^* \cdot \tilde \eta^*)(p\cdot q)-(\varepsilon^* \cdot p)(\tilde \eta^* \cdot q) \right]   \nn \label{eq:ampDs0} \\
\langle \gamma(q,\lambda) D_s(p)| D_{s1}^\prime(p+q,\lambda^{\prime\prime})\rangle &=&  e g_1  \left[ (\varepsilon^* \cdot  \eta)(p \cdot q)-(\varepsilon^* \cdot p)(\eta \cdot q) \right]   \nn \label{eq:ampDs1a} \\
\langle \gamma(q,\lambda) D^*_s(p,\lambda^\prime)| D_{s1}^\prime(p+q,\lambda^{\prime \prime})\rangle &=& i \, e \, g_2 \,  \varepsilon_{\alpha \beta \sigma \tau}  \eta^\alpha \tilde \eta^{*\beta} \varepsilon^{*\sigma}   q^\tau  \label{eq:ampDs1b} \\
\langle \gamma(q,\lambda) D_{s0} (p)| D^\prime_{s1}(p+q,\lambda^{\prime \prime})\rangle &=& i \, e \,  g_3  \, \varepsilon_{\alpha \beta \sigma \tau}  \varepsilon^{*\alpha} \eta^\beta    p^\sigma q^\tau \nn
\label{eq:ampDs1Ds0}
\eea
($\varepsilon(\lambda)$ is  the photon 
 polarization vector and $\tilde \eta(\lambda^\prime)$,   
$\eta(\lambda^{\prime\prime} )$ the $D_s^*$  and $D_{s1}^\prime$ polarization vectors).
Considering   the correlation functions\cite{altri,Colangelo:2000dp}
\be
F(p,q)=i \int d^4x \; e^{i p \cdot x} \langle \gamma(q,\lambda) | T[J^\dagger_A(x) J_B(0)] |0\rangle
\label{eq:corr-Ds0Ds*gamma}
\ee
of  quark-antiquark currents  $J_{A,B}$ having the same quantum number of the decaying and of the produced charmed mesons,  and an external photon state of momentum $q$ and helicity $\lambda$,
%
%%%%%%%%%%%%%%%%%%%%%%%%%%%%%%%%%%%%%%%%%%%%%%%%
\begin{figure}[h]
 \begin{center}
\includegraphics[width=0.8\textwidth] {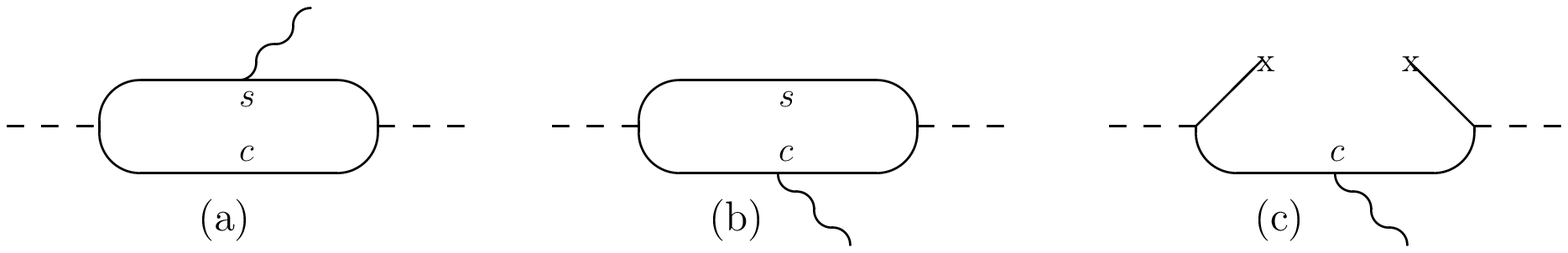}\\
\includegraphics[width=0.6\textwidth] {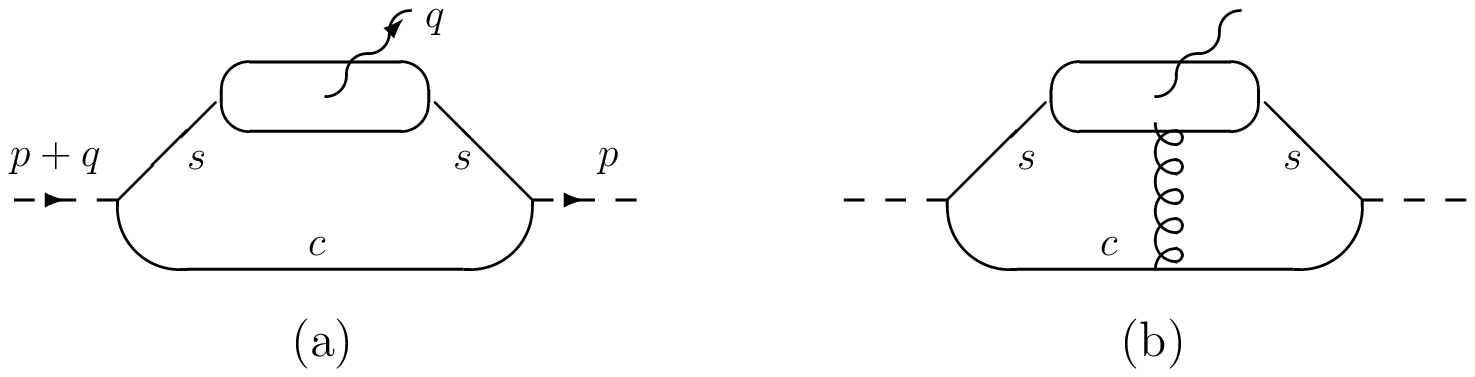}\\
\vspace*{0mm}
 \caption{ Leading contributions to the correlation functions eq.(\ref{eq:corr-Ds0Ds*gamma})
 expanded on the light-cone:
 perturbative photon emission by the strange  and  charm  quark ((a,b) in the first line)
 and  two- and three-particle photon distribution amplitudes (second line);
 (c)  corresponds to the strange quark condensate contribution.}
  \label{fig:light-cone}
 \end{center}
\end{figure}
%%%%%%%%%%%%%%%%%%%%%%%%%%%%%%%%%%%%%%%%%%%%%%%%%
%
and expanding on the light-cone, it is possible to express $F$ in terms of
the perturbative  photon coupling to the strange and charm quarks,
 together with  the contributions of the photon emission from the soft $s$ quark, 
expressed as photon matrix elements of increasing twist\cite{Ball:2002ps}, see fig.\ref{fig:light-cone} . The hadronic representation of the correlation function involves the contribution of the lowest-lying resonances,  the current-vacuum matrix elements of which 
are computed by the same method\cite{Colangelo:1995ph}, 
 and a continuum of states  treated invoking  global quark-hadron duality.
 A Borel transformation  introduces an external parameter $M^2$, 
  the hadronic quantities being
 independent of it (fig. \ref{fig:results}).  
%
%%%%%%%%%%%%%%%%%%%%%%%%%%%%%%%%%%%%%%%%%%%%
\begin{table}[b]
\tbl{
Radiative decay widths (in keV) of $D^*_{sJ}(2317)$ and $D_{sJ}(2460)$ obtained by Light-Cone sum rules
(LCSR),    Vector Meson Dominance (VMD) and constituent quark model (QM).} 
{\begin{tabular}{c c c c c c}
\hline Initial state & Final state & LCSR  \cite{Colangelo:2005hv}&VMD \cite{colangelo1}
& QM \cite{Godfrey:2003kg}  & QM \cite{Bardeen} \\ \hline 
$D^*_{sJ}(2317)$&$D_{s}^{\ast }\gamma$&  4-6     & 0.85& 1.9 &  1.74\\ 
$D_{sJ}(2460)$   &$D_{s}\gamma$             & 19-29 & 3.3&  6.2 & 5.08 \\
                              &$D_{s}^*\gamma$          & 0.6-1.1 & 1.5&  5.5 & 4.66 \\
                               &\,\,\,\,\,$D^*_{sJ}(2317)\gamma$\,\,\,\,\,& 0.5-0.8  &  \ \hfill --- \hfill\ &0.012 & 2.74\\ \hline
\end{tabular}}\label{predictions}
\end{table}
%%%%%%%%%%%%%%%%%%%%%%%%%%%%%%%%%%%%%%%%%%%%

Looking at the results, collected in Table \ref{predictions},
one sees   that the rate of $D_{sJ}(2460) \to D_{s}\gamma$  is the largest one among the radiative  $D_{sJ}(2460)$ rates, and this is confirmed by 
experiment, as reported in Table \ref{br}.  
%
%%%%%%%%%%%%%%%%%%%%%%%%%%%%%%%%%%%%%%%%%%%%
\begin{table}[t]
\tbl{Measurements and 90\% CL  limits of  ratios of  $D^*_{sJ}(2317)$  and  $D_{sJ}(2460)$ decay widths.} 
{\begin{tabular}{cccc}
\hline & Belle & BaBar   & CLEO  \\ \hline $\Gamma \left(
D^*_{sJ}(2317) \rightarrow D_{s}^{\ast }\gamma \right)/ \Gamma \left(
D^*_{sJ}(2317)\rightarrow D_{s}\pi ^{0}\right) $ & $<0.18$&  \ \hfill --- \hfill\   & $<0.059$ \\
$\Gamma \left(D_{sJ}(2460) \rightarrow
D_{s}\gamma \right) /\Gamma \left( D_{sJ}(2460)\rightarrow
D_{s}^{\ast }\pi ^{0}\right) $ & $0.45 \pm 0.09$ & $0.30 \pm 0.04$ & $<0.49$ \\
$\Gamma \left( D_{sJ}(2460)\rightarrow
D_{s}^{\ast }\gamma \right) /\Gamma \left( D_{sJ}(2460)\rightarrow D_{s}^{\ast }\pi ^{0}\right) $ & $<0.31$ &\ \hfill --- \hfill\ &$<0.16$ \\
$\Gamma \left( D_{sJ}(2460)\rightarrow
D^*_{sJ}(2317)\gamma \right)/ \Gamma \left( D_{sJ}(2460)\rightarrow
D_{s}^*\pi^0 \right) $  & \ \hfill --- \hfill\ & $ < 0.23$  & $ < 0.58$\\
\hline
\end{tabular}}\label{br}
\end{table}
%
%%%%%%%%%%%%%%%%%%%%%%%%%%%%%%%%%%%%%%%%%%%%
%
Quantitative understanding of  the  ratios in Table \ref{br} requires a  precise knowledge of
the  widths of the isospin violating transitions
$D_{s0}\to D_s \pi^0$ and $D^\prime_{s1}\to D_s^* \pi^0$. In the description of    
these  transitions based on the mechanism of $\eta-\pi^0$ mixing \cite{Godfrey:2003kg,colangelo1} the accurate determination of  the
strong   $D_{s0} D_s \eta$ and $D^\prime_{s0} D^*_s \eta$  couplings
for finite heavy quark mass and including $SU(3)$ corrections is required.   

If there are no reasons to consider $D^*_{sJ}(2317)$ and $D_{sJ}(2460)$ as exotic
mesons, the same conclusion seems mandatory for $D_{sJ}(2860)$, a
 state recently observed by  BaBar\cite{palano06}   in the $DK$ system inclusively produced in $e^+ e^- \to DKX$.  The  parameters of the  resonance  are:
$M(D_{sJ}(2860))=2856.6 \pm 1.5 \pm 5.0$ MeV  and
$\Gamma(D_{sJ}(2860) \to DK)= 48\pm 7 \pm10$  MeV,
(where $DK=D^0 K^+$ and $D^+ K_S$). In the same set of data and range of mass no structures seem to appear in the $D^*K$  distribution, while
a broad contribution seems to be present  in the $DK$ 
distribution at  smaller mass. 
 
 It is interesting to discuss this new meson in some detail\cite{nicotri06}.
A possible quantum number assignment  for a $c \bar s$ meson decaying to $DK$  is either  
$s_\ell^P=\frac{3}{2}^-$ $J^P=1^-$,   or 
$s_\ell^P=\frac{5}{2}^-$ $J^P=3^-$,  in both cases  corresponding
to  $\ell=2$ and  lowest radial quantum number ($n=0$).  
Another possibility is that $D_{sJ}(2860)$ is a radial excitation ($n=1$) of already observed $c\bar s$ mesons:  the  $J^P=1^-$   $ s_\ell^P=\frac{1}{2}^-$ state (the first radial excitation of $D_s^*$),  
the  $J^P=0^+$   $ s_\ell^P=\frac{1}{2}^+$ state (radial excitation of $D_{sJ}^*(2317)$) or
the   $J^P=2^+$   $ s_\ell^P=\frac{3}{2}^+$ state (radial excitation of $D_{s2}(2573)$). 
In the absence of the helicity distribution of the final state,  arguments can be provided to  support a particular assignment of $J^P$  considering the observed mass,  the decay modes and width.
 
A piece of information  comes from the  $DK$ width. Using an effective QCD Lagrangian
incorporating spin-flavour heavy quark symmetry and light quark chiral symmetry, an  estimate is possible of   the ratios 
$ \displaystyle \frac{\Gamma( D_{sJ} (2860) \to D^*K) }{\Gamma( D_{sJ} (2860)\to DK) }$ 
and  $\displaystyle \frac{\Gamma( D_{sJ} (2860)\to D_s \eta}{\Gamma( D_{sJ} (2860)\to DK)  }$
 for various quantum number assignments to $D_{sJ} (2860)$
(Table \ref{ratios}).
\begin{table}[b]
\tbl{Predicted $ \displaystyle \frac{\Gamma( D_{sJ} \to D^*K)}{\Gamma( D_{sJ} \to DK)}$  and   
$\displaystyle \frac{\Gamma( D_{sJ} \to D_s \eta)}{\Gamma( D_{sJ} \to DK)  }$   for  various assignment  of quantum numbers to  $D_{sJ}(2860)$. The sum $DK=D^0K^+ + D^+ K_S$ is understood.}
{\begin{tabular}{ c  c  c  c }
      \hline
 $s_\ell^p$, $J^P$,  $n$   &  $D_{sJ}(2860) \to DK $&$\displaystyle\frac{\Gamma( D_{sJ} \to D^*K)}{\Gamma( D_{sJ} \to DK)
 }$  &   $\displaystyle\frac{\Gamma( D_{sJ} \to D_s \eta)}{ \Gamma( D_{sJ} \to DK)  }$ 
\\ \hline
 $\frac{1}{2}^-$, $1^-$,  $1$ & $p$-wave &$1.23$& $0.27$ \\
$\frac{1}{2}^+$, $0^+$, $1$&  $s$-wave &$0$& $0.34$ \\
$\frac{3}{2}^+$, $2^+$, $1$&  $d$-wave &$0.63$& $0.19$\\
$\frac{3}{2}^-$,  $1^-$,  $0$ & $p$-wave  & $0.06$& $0.23$ \\
$\frac{5}{2}^-$,  $3^-$,  $0$ & $f$-wave  & $0.39$& $0.13$ \\
\hline
\end{tabular}}\label{ratios}
\end{table}
Non observation (at present)  of a $D^*K$ signal implies that
the production of $D^* K$ is not favoured, and therefore  the assignments
$s_\ell^P=\frac{1}{ 2}^-$, $J^P=1^-$,  $n=1$,  and $s_\ell^P=\frac{3}{ 2}^+$, $J^P=2^+$, $n=1$
can be excluded.  The assignment 
$s_\ell^P=\frac{3}{2}^-$, $J^P=1^-$, $n=0$ can also be excluded, since   the  width  $ \Gamma(D_{sJ}\to DK)$ would be naturally large  for a  $p-$wave  $D_{sJ} \to DK$ transition.
\footnote{A  candidate for this assignment is  the resonance $D_{sJ}(2715)$   observed  very recently by Belle\cite{belle2715} 
in $B^+\to \bar D^0 D^0 K^+$ decays with  $M=2715\pm11^{+11}_{-14}$ MeV,    $\Gamma=115\pm20^{+36}_{-32}$ MeV
and $J^P=1^-$,  a state that could also be interpreted as the first radial recurrence of $D_s^*$, as discussed in Refs.~\refcite{close,Zhang:2006yj}.}

In the case of the assignment  $s_\ell^P=\frac{1}{2}^+$, $J^P=0^+$, $n=1$ 
the decay  $D_{sJ}\to D^*K$ is forbidden;  this is the assignment proposed in Refs.~\refcite{vanBeveren:2006st} and~\refcite{close}.
However, $D_{sJ}\to DK$  occurs  in $s-$wave, therefore it should be rather broad: 
for the state with the lowest radial quantum
number $n=0$ the computed coupling costant $g_{D_{sJ}DK}$  is  in agreement  with 
observation\cite{Colangelo:1995ph,Colangelo:2003vg}, and using it  one would obtain  $\Gamma(D_{sJ}\to DK)\simeq 1.4$ GeV.  Although  it is reasonable to suppose that the coupling of radial excitation is smaller,  the  suppression should be substantial  to reproduce the observed width.
Moreover, a large signal would be expected  in the $D_s \eta$ channel. Another remark is
that the spin partner with $J^P=1^+$ ($s_\ell^p=\frac{1}{2}^+$, $n=1$) 
would decay to $D^* K$ with a small width, $\simeq 40$ MeV, a rather easy signal to observe;
therefore,  to explain the absence of the $D^*K$ signal one must invoke a mechanism favouring the production of the $0^+$ state and inhibiting that of $1^+$ state in 
$e^+ e^- \to DKX$, a mechanism discriminating the first radial excitation from the case $n=0$.

For  $s_\ell^P=\frac{5}{2}^-$, $J^P=3^-$, $n=0$
 the small $DK$ width is mainly due to kinematics ($\Gamma \propto q_K^7$).
A smaller but non negligible signal
in the $D^*K$ mode is predicted,  and  a small signal in the $D_s \eta$ mode is also expected.
 The coupling constant is  similar to the  couplings of the other doublets to light
pseudoscalars. If  $D_{sJ}(2860)$ has $J^P=3^-$, it is not expected to be  produced
 in non leptonic $B$ decays such as  $\bar B^0 \to D_{sJ}(2860)^- D^+$
 and  $B^- \to  \bar D_{sJ}(2860)^-  D^0$,  so that
 the quantum number assignment can be confirmed by studies of $D_{sJ}$ production in $B$
transitions.   $D_{sJ}(2860)$ can be  one of the predicted high mass, high spin and relatively narrow $c \bar s$ states   \cite{Colangelo:2000jq,nicotri06};  its non-strange partner  $D_3$ is  also expected to be narrow: $\Gamma(D_{3}^+\to D^0 \pi^+ )\simeq 37$ MeV, and  can  be 
produced in semileptonic and  non leptonic $B$ decays, 
such as $\bar B^0 \to D_3^+ \ell^-  \bar \nu_\ell$ and $\bar B^0 \to D_3^+ \pi^-$.   

\begin{figure}[hbt]
\vspace*{-12mm}
 \begin{center}
\includegraphics[width=0.90\textwidth] {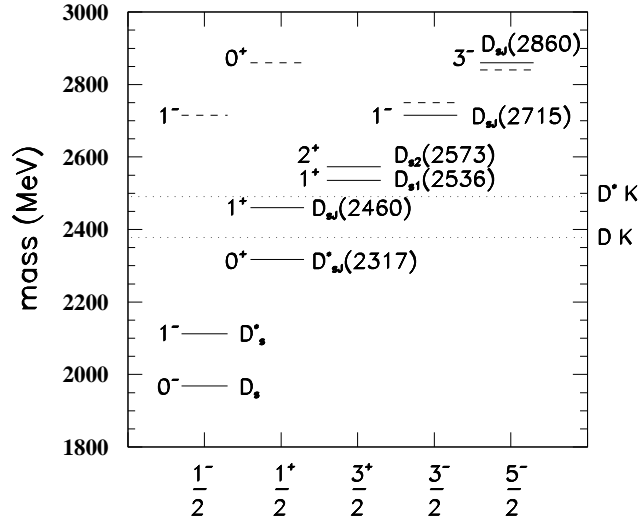}
\vspace*{-12mm}
 \caption{Possible classification  of the known $c \bar s$ states according to $s_\ell^P$. The positions
 of $D_{sJ}(2869)$ and $D_{sJ}(2715)$ if interpreted as radial excitations   are also shown.}
  \label{doublets}
 \end{center}
\end{figure}

We conclude this Section showing   in fig. \ref{doublets}  a  tentative classification of  the known $c \bar s$ mesons. 
 Confirmation of this classification and the search for the missing states is a task for current and future investigations.

%%%%%%%%%%%%%%%%%%%%%%%%%%%%%%%%%%%%%%%%%%%%%%%%%
%%%%%%%%%%%%%%%%%%%%%%%%%%%%%%%%%%%%%%%%%%%%%%%%%
 
 \section{Hidden charm mesons}\label{sec:hiddenc}

While  the results  in the open charm sector can be organized in a well-established scheme, 
the situation  in the hidden charm sector is more complex. A few new results, in particular those concerning $\eta^\prime_c$ and $h_c$,  essentially agree with the expectations,
although some particular aspects deserve  investigations. Others, namely those concerning 
$X(3940)$, $Y(3940)$ and $Z(3930)$, could be organized according to generally accepted schemes
with some caveat. The observations concerning $Y(4260)$ and $X(3872)$ have
puzzling aspects:  in particular, these states present features that could be expected for non standard quark-antiquark mesons, as we  briefly discuss below. 

\subsection{$h_c$ and $\eta_c^\prime$}
The observation of   $h_c$ ($J^{PC}=1^{+-}$)  by Cleo\cite{hc-Cleo}  in 
$\psi^\prime \to \pi^0 h_c$, with $h_c$ decaying to $\eta_c \gamma$, completes
the set of four  low-lying charmonium states with $\ell=1$. The mass:
$M(h_c)=3524.4\pm0.6\pm0.4$ MeV deviates by less than $1$ MeV from the center of gravity of the $\chi_{cJ}$ states. The strategy of searching $h_c$ in $B$ decays\cite{Colangelo:2003sa} has not been successful, yet,  since the branching fraction of $B \to K h_c$ is smaller than estimated 
by the  methods that reproduce  the measured $B(B \to K \chi_{c0})$  \cite{Fang:2006bz}.

Also the observation of  $\eta_c^\prime$, made by Belle, Cleo and BaBar in $B$ decays: $B \to K \eta_c^\prime$, in $e^+e^- \to J/\psi \eta_c^\prime$ and in $\gamma \gamma \to \eta_c^\prime \to K_S K^\pm \pi^\mp$ \cite{etacprime}  completes the doublet of the first radial excitations of $(\eta_c, J/\psi)$. The parameters of the resonance are: $M(\eta^\prime_c)=3638 \pm 4$ MeV (thus the hyperfine splitting is $48$ MeV)
and $\Gamma(\eta^\prime_c)=14 \pm 7$ MeV.  The observations are  in agreement with the
expectations, with some difficulty with the $\gamma \gamma$ rate of $\eta_c^\prime$ 
which is smaller  than estimated\cite{pham}. The $c \bar c$ spectrum below the open charm threshold can be reproduced by a one-gluon-exchange short-distance potential, a scalar linearly confining potential and spin-spin and spin-orbit interactions\cite{barnes06}. However, when the energy increases, the theoretical
determination of the meson properties, in particular of the spectrum,  cannot ignore the  open charm thresholds, starting from $D^0 \bar D^0$,  an old problem for which there is no
model-independent solution, yet.  Mass shifts of   $20-40$  MeV have been estimated
for states close to the thresholds \cite{eichten}.  These effects must be considered 
in the discussion of  $X(3940)$, $Y(3940)$ and $Z(3930)$.
\subsection{$X(3940)$, $Y(3940)$ and  $Z(3930)$.}
For $X(3940)$, found by Belle\cite{X3940}  in the hadronic system recoiling against $J/\psi$ in $e^+e^-$ annihilation,
with  $M=3943\pm6\pm6$ MeV, $\Gamma<52$ MeV and decays
into $D^* \bar D$,  two interpretations are possible:  i) the $3^1S_0$ partner of  $3^3S_1$
($\psi(4040)$), an assignment that could be confirmed by observation of the state in $\gamma \gamma$;
ii) the first radial excitation of $\chi_{c1}$,  with the difficulty that  $\chi_{c1}$  has not been found in the same 
set of data; moreover,   another candidate, $Y(3940)$, is available for the same assignment.

Indeed, $Y(3940)$ was also found by Belle\cite{Y3940} in the  $J/\psi \omega$ system produced in $B \to K  J/\psi \omega$. 
Its parameters are: $M=3943\pm11\pm13$ MeV and $\Gamma=85\pm22\pm26$ MeV;  decays
 to open charm mesons have not been found, so far. The possible assignment as  $2^3P_1$ ($\chi_{c1}^\prime$) implies that it should  be observed in $DD^*$, even though the phase space for such a mode is small.

$Z(3930)$ is the last state in this region of mass found by Belle\cite{Z3930} in $\gamma \gamma \to D \bar D$, with
$M=3941\pm 4\pm 2$ MeV and $\Gamma=20 \pm 8\pm 3$ MeV.
  The helicity distribution in the final state is consistent with a $J=2$ state,  therefore it can be 
  identified as the  $2^3P_2$ ($\chi_{c2}^\prime$) meson. 

\begin{figure}[h]
\vspace*{-15mm}
 \begin{center}
\includegraphics[width=0.90\textwidth] {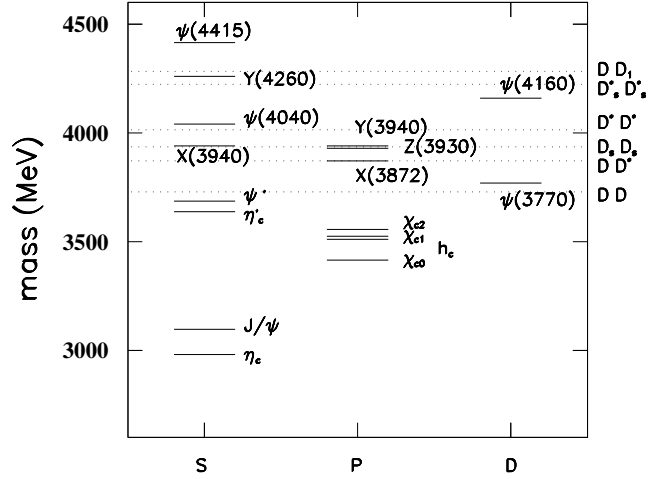}
\vspace*{-15mm}
 \caption{Spectrum of $c \bar c$ states together with the thresholds for decays to open charm mesons.
 Possible positions of $X(3872)$ and $Y(4260)$ are shown.}
  \label{fig:spectrumcc}
 \end{center}
\end{figure}

In spite of the uncertainties in the quantum number assignment,  the three states  can be arranged in the  $c \bar c$ spectrum, as shown in fig. \ref{fig:spectrumcc}.
The case of $Y(4260)$ and $X(3872)$ is more difficult.
%
%%%%%%%%%%%%%%%%%%%%%%%%%%%%%%%%%%%%%%%%%%%%%%%%%
%%%%%%%%%%%%%%%%%%%%%%%%%%%%%%%%%%%%%%%%%%%%%%%%%
%
\subsection{$Y(4260)$}
$Y(4260)$ is the first  meson in  the list of states seeming to escape ordinary classifications.  It
was found by BaBar\cite{Y4260-BaBar} in $B^- \to K^- \pi \pi J/\psi$ and 
in radiative return analyses $e^+ e^- \to \gamma_{ISR} \pi \pi J/\psi$,
and confirmed by Cleo\cite{Y4260-Cleo} in $e^+ e^- \to \gamma_{ISR}  Y$, with $Y$
observed in $\pi^+ \pi^- J/\psi$, $\pi^0 \pi^0 J/\psi$ and $K^+ K^- J/\psi$.
The properties of the resonance are:
$M=4259\pm8\pm4$ MeV, $\Gamma=88\pm23\pm5$ MeV and $J^{PC}=1^{--}$.
Moreover, the dipion mass distribution is consistent with a $s$-wave structure, so that a decay through $f_0(980)$  can be supposed.
The problem with a  $c \bar c$ interpretation  is that a $1^{--}$ meson can be
either a $\ell=0$ state, a radial excitation between $\psi(4040)$ ($\psi(3S)$)
and $\psi(4415)$ (at present  interpreted as $\psi(4S)$), or a $\ell=2$ state above
$\psi(4159)$ (interpreted as $\psi(2D)$), with mass  not predicted by any theoretical determination. Therefore, the meson looks as an extra state with respect to the $1^{--}$ levels,
a state with a large coupling to $\pi \pi J/\psi$ and without  observed  (so far) decays in open charm mesons. Its mass  is just above the $\bar D_s^*  D_s^*$ threshold and below the  $\bar D D_1$ threshold, $D_1(2420)$ being the narrow  $c \bar q$ axial vector state. Among various  interpretations\cite{interpretations-Y},   
the one suggesting that $Y(4260)$ is a $\bar c G c$  $1^{--}$
hybrid\cite{pene} emphasizes the agreement of the observation with some expectations. Indeed,
 charmed hybrids in this
 range of mass are conjectured, namely on the basis of  lattice QCD simulations, with large couplings to 
$J/\psi$ and light  ($\eta, \eta^\prime$) mesons and  with  decays in open charm mesons with different orbital angular momentum (a decay in $D D_1^\prime(2440)$ is possible,
due to the broad width of $D_1^\prime$). Noticeably, other hybrids with different quantum numbers are extected in the same range of mass;  their observation would open a new chapter of the  hadron spectroscopy.

\subsection{$X(3872)$}
We have left  $X(3872)$ as the last meson to discuss, since it presents the most puzzling aspects.
The observations   can be summarized as  follows:
\begin{romanlist}[]
\item the $X$  resonance has been  found in $J/\psi \pi^+ \pi^-$ distribution  by four experiments, both in $B$ decays
($B^{-(0)} \to K^{-(0)} X$), both
in $p \bar p$ annihilation\cite{expX3872}. The  mass is $M=3871.9 \pm 0.6$ MeV while the width remains unresolved:  $\Gamma < 2.3$ MeV (90 \% CL);
\item  there is no evidence of resonances  in the charged mode $J/\psi \pi^\pm \pi^0$ or in $J/\psi \eta$ \cite{babarkpi};
\item the state is not observed in $e^+ e^-$ annihilation;
\item  for $X$ produced in $B$ decays the ratio $\displaystyle \frac{B(B^0 \to K^0 X)}{B(B^+ \to K^+ X)}=0.61 \pm 0.36 \pm 0.06$ is obtained\cite{babarkpi};
\item the dipion spectrum in $J/\psi \pi^+ \pi^-$ is peaked at large mass;
\item  the decay  in $J/\psi \pi^+ \pi^- \pi^0$  is observed\cite{belle3p} with  $\displaystyle \frac{B(X \to J/\psi \pi^+ \pi^- \pi^0)}{B(X \to J/\psi \pi^+ \pi^- )}=1.0 \pm 0.4 \pm 0.3$: this implies G-parity violation; 
\item the radiative  mode $X \to J/\psi \gamma$ is found\cite{belle3p,babarpsigamma}  with 
$\displaystyle \frac{B(X \to J/\psi \gamma)}{B(X \to J/\psi \pi^+ \pi^- )}=0.19 \pm 0.07$, therefore charge conjugation of the state is C=+1;
\item  the angular distribution of the final state is compatible with the spin-parity assignment $J^P=1^+$  \cite{Abe:2005iy};
\item  there is a  signal in $D^0 \bar D^0 \pi^0$ with  $\frac{B(X \to D^0 \bar D^0  \pi^0)}{B(X \to J/\psi \pi^+ \pi^- )}=9\pm4$   \cite{BelleDoDopi} .
\end{romanlist} 

All the measurements are thus compatible with the assignment $J^{PC}=1^{++}$.  
If the $2^3P_2$ is identified with  $Y(3940)$,  there is  overpopulation of $1^{++}$ $c \bar c$ mesons.

Noticeably,  the mass of the resonance coincides with that of the $D^{*0} \bar D^0$ pair;  this suggests  that the state could be  a realization of the molecular quarkonium\cite{okun}, a bound state of two mesons, $D^{*0}$ and $\bar D^0$,
with small binding energy\cite{interpretations}.
 The absence of a   $D^{*+}  D^-$ molecule
can be interpreted in this scheme observing that, being  heavier by 7 MeV, such a state can 
rapidly decay in $D^{*0} \bar D^0$. 
In this description, the wave function of X(3872) has various components\cite{voloshin}:
\be
|X(3872)>=a \, |D^{*0} \bar D^0+ \bar D^{*0}  D^0> + b \,  |D^{*+}  D^-+  D^{*-}  D^+> + \dots
\ee
allowing to explain a few  observations and to make predictions:
\begin{romanlist}[]
\item the state has  no definite isospin;
\item the decay $X \to J/\psi \pi^0 \pi^0$ is forbidden; 
\item since the decays of the resonance are mainly due to the decays of its components, the radiative
transition in neutral mesons $X \to D^0 \bar D^0 \gamma$ should be dominant with respect to $X \to D^+  D^- \gamma$;
\item a  resonance $X_b(10604)$ is expected as a bound state of $\bar B^0$ and $B^{*0}$;
\item if the molecular binding mechanism is provided by a single pion exchange, this model explains the absence of  $D \bar D$ molecular states.
\end{romanlist}
The description of $X(3872)$ in the simple charmonium  scheme, leaving unsolved  the issue of the overpopulation of $1^{++}$ states,   presents alternative arguments  to the molecular description\cite{suzuki}.  First,   the molecular binding mechanism  cannot be  a single
$\pi^0$ exchange, since this produces an attractive  potential which  is a delta function in space:
\be
V(r)=-\frac{1}{3} \, g^2_{D^*D\pi} \, \epsilon^\prime \cdot \epsilon \, \delta(r)+\dots
\ee
($g_{D^*D\pi}$ is the coupling constant of the $D^* D \pi$ vertex, $\epsilon$ and $\epsilon^\prime$ 
the $D^*$ polarization vectors)
and therefore it does not give rise to a bound state in three spatial dimensions. 
Concerning the isospin (G-parity) violation,  to correctly  interpret the large value of the ratio
$\displaystyle \frac{B(X \to J/\psi \pi^+ \pi^- \pi^0)}{B(X \to J/\psi \pi^+ \pi^- )}$ one has to consider that
the phase space effects in two and three pion modes are very different. 
The amplitude ratio is  rather small:
$\displaystyle \frac{A(X \to J/\psi \rho^0)}{A(X \to J/\psi \omega)}\simeq 0.2$, so that the isospin violating amplitude is
20\% of the isospin conserving one, an 
 effect that could be related to another isospin violating effect,  the mass difference between 
neutral and charged $D$ mesons, considering  the contribution of  $DD^*$ intermediate states
to $X$ decays. Finally,  also the eventual  dominance of
$X \to D^0 \bar D^0 \gamma$ with respect to $X \to D^+  D^- \gamma$ could be interpreted invoking
standard mechanisms. Notice that
a prediction  of the charmonium description is that the rates of 
$B^0 \to X K^0$ and $B^- \to X K^-$ are nearly equal;   the measurements  are   not conclusive on this point. 

Our conclusion is that, at present,  there are no compelling arguments allowing to exclude an interpretation in favour of  others. Further analyses are requested to solve  the issue of $X(3872)$.

\section{Conclusions}\label{sec:concl}
In this short review of the new charm meson spectroscopy we have attempted to schematically describe the experimental observations, various interpretations and the main open problems.
We do not want to emphasize how interesting  the present situation is, and how much work is
needed, both on the experimental, both on the theory side, to elaborate the   information 
collected so far. We prefer to borrow the conclusion from another review on charm, written
about 30 years ago:
"It is easy to see the time when the charmed particles will be studied in detail... so that we look for
new enjoyment and surprises." \cite{shifman}

\vspace*{5mm}

\noindent{\bf Acknowledgments}

\vspace*{3mm}

\noindent
We are grateful to  M.A.Shifman  for  invitation to the QCD Workshop. We thank
 A. Palano and T.N. Pham for discussions.
%, and we acknowledge partial support from the EC Contract No. HPRN-CT-2002-00311 (EURIDICE).


\begin{thebibliography}{9}
\bibitem{interpretations}
Other proposals are discussed in Ref. ~\refcite{swanson}.
\end{thebibliography}
\end{document}